\documentclass[aps, showkeys, showpacs]{revtex4}
\usepackage[]{graphicx}
\usepackage[]{amsmath, amsfonts, bm}
\begin{document}

\title{Heat capacity and phonon mean free path of wurtzite GaN}
\author{B A Danilchenko\dag \footnote{Author to whom correspondence should be addressed; electronic mail: danil@iop.kiev.ua}, T Paszkiewicz\ddag, S Wolski\ddag, A. Je{\.z}owski\S, and T.Plackowski\S} 

\begin{abstract}
We report on lattice specific heat of bulk hexagonal GaN measured by the heat flow method in the temperature range 20-300 K and by the adiabatic method in the range 5-70 K. We fit the experimental data using two temperatures model. The best fit with the accuracy of 3 $\%$ was obtained for the temperature independent Debye's temperature  $\theta_{\rm D}=365$ {\rm K} and Einstein's temperature  $\theta_{\rm E}=880$ {\rm K}. We relate these temperatures to the function of density of states. Using our results for heat conduction coefficient, we established in temperature range 10-100 K the explicit dependence of the phonon mean free path on temperature  $\it{l}_{\rm ph}\propto T^{-2}$. Above 100 K, there is the evidence of contribution of the Umklapp processes which limit phonon free path at high temepratures. For phonons with energy $k_{\rm B}\times 300 $ {\rm K} the mean free path is of the order 100 {\rm nm}. 
\end{abstract}
\affiliation{\dag\
Institute of Physics, National Academy of Sciences of Ukraine,
Prospect Nauki 46, 252650 Kiev, Ukraine} \affiliation{\ddag\ Chair
of Physics, Rzesz\'{o}w University of Technology, ul. W. Pola 2,
PL-35-959 Rzesz\'{o}w Poland} 
\affiliation{\S Institute of Low Temperature and Structure Reasearch, Polish Academy of Sciences, 50-422 Wroc{\l}aw, ul. Ok\'{o}lna 2}
\maketitle

\section{Introduction}
\label{sc:introduction}
High pressure grown GaN has been used as a substrate for high quality epitaxial layers, heterostructures and blue laser diode \cite{1}. Efficient heat removal is critical to the performance of high power semiconductor electronic devices, such as light emitting diodes, lasers, photodetectors or metal-semiconductor field effect transistors, and high electron mobility transistors. For this reason, thermal conductivity and heat capacity are mandatory characteristics of GaN, a very topical semiconductor these days. In our previous paper we studied thermal conductivity of this compound \cite{2}. 

Early measurements of the heat capacity of GaN had been performed for polycrystalline powder samples \cite{3}. Measurements of the heat capacity of hexagonal single crystals of GaN in the 20-1400 K temperature range were reported by Kremer et al \cite{4} and for 320-570 {\rm K} range for powder samples by Leitner et al \cite{5}. Their results match findings of Kashchenko et al for powder samples \cite{3} very well. 

In our paper we focus our attention on the bulk wurzite GaN heat capacity $C_{\rm p}$ measured in the low temperature region. Having measured $C_{\rm p}$ and thermal conductivity coefficient $\kappa$ \cite{2}, we are able to estimate the temperature dependence of the mean free path $l_{\rm ph}$ of acoustic phonons. The knowledge of $\it{l}_{\rm ph}$ is important for understanding the thermalization processes in GaN substrates and structures as well as for the heat thermal budget calculations.  	

The phonon density of states (DOS) $g(\omega)$, which includes contributions from all phonons over the entire Brillouin zone, is needed for the calculations of various thermodynamic characteristics, e.g. heat capacity and thermal conductivity. As neutrons probe phonon modes throughout the Brillouin zone, the neutron scattering can be applied to establish DOS. Nipko et al experimentally established DOS for powder samples of GaN \cite{6}. The partial DOS of near zone center optical phonons can be also obtained using Raman and infrared spectroscopy (in the case of GaN cf. references given in \cite{6}). DOS of GaN was calculated by Nipko et al \cite{6}, Azuhata et al \cite{7}, Bungaro et al \cite{8} and  Sanati and Estreicher \cite{9}. 

Debye proposed a simple form of DOS based on the elastic continuum theory 
\begin{equation}
g_{\rm D}(\omega)=3\frac{{\omega}^{2}}{\omega_{\rm D}^{3}}\theta(\omega_{\rm D}-\omega), 
\label{Debye-DOS}
\end{equation}
where $\theta(x)$ is the Heaviside step function. $g_{\rm D}(\omega)$ depends only on one parameter -- the Debye frequency  $\omega_{\rm D}$ (or on the Debye temperature $\theta_{\rm D}=\hbar \omega_{\rm D}/k_{\rm B}$). The Einstein form of DOS 
\begin{equation}
	g_{\rm E}=3Ns\delta (\omega-\omega_{\rm E})
\label{Einstein-DOS}
\end{equation}
where  $\delta(x)$ is Dirac's delta distribution, N -- the number of elementary cells, and s -- the number of particles in it, assumes that all lattice particles vibrate with the same frequency of optic phonons, $\omega_{\rm E}$, hence the DOS depends also on the Einstein frequency $\omega_{\rm E}$ or on the corresponding temperature $\theta_{\rm E}=\hbar \omega_{\rm E}/k_{\rm B}$. Debye's and Einstein's models with several characteristic temperatures (frequencies) were also introduced (cf. \cite{ 10}). 
	
In our paper we discuss the results of measurements of the wurzite GaN heat capacity in terms of a model depending on the Debye and Einstein temperatures \cite{10}, namely we assume that $C_{\rm p}$ $\rm [J/mol$ $\rm K]$
\begin{eqnarray}
\nonumber &C_{\rm p}(T)\approx C_{\rm D}(T)+C_{\rm E}(T)=\\
&3N_{\rm A}k_{\rm B}3\left(T/\theta_{\rm D}\right)^{3}F(T/\theta_{\rm D})+\\
\nonumber &+3N_{\rm A}k_{\rm B}\left(\theta_{\rm E}/T\right)^{2}e^{\theta_{\rm E}/T}/\left(e^{\theta_{\rm E}/T}-1\right)^{2},
\label{eq:total-c_p}
\end{eqnarray}
where 	
\begin{equation}
	F(T/\theta_{\rm D})=\int_{0}^{T/\theta_{\rm D}}dx\,x^{4}e^{x}/\left(e^{x}-1\right)^{2},
\label{eq:Debye-function}
\end{equation}
and $N_{\rm A}$ is the Avogadro number.

Hexagonal GaN single crystals were grown at the Institute of High Pressure Physics, Polish Academy of Sciences \cite{11} and in ETH Z{\"u}rich. The specific heat of approximately 30 {\rm mg} GaN single crystal was measured using the heat-flow calorimeter \cite{12}. In this method, $C_p$ measurements could be performed both upon cooling and heating. The high data density is another advantage of this method (approximately 10 points/{\rm K}). Using this method we measured heat capacity in the temperature range 20-300 K. Adiabatic heat capacity measurements were performed in the temperature range 5-70 {\rm K} on three different samples with masses approximately 100 {\rm mg}. 

\begin{figure}[tbp]
\begin{center}
\includegraphics[width=7cm]{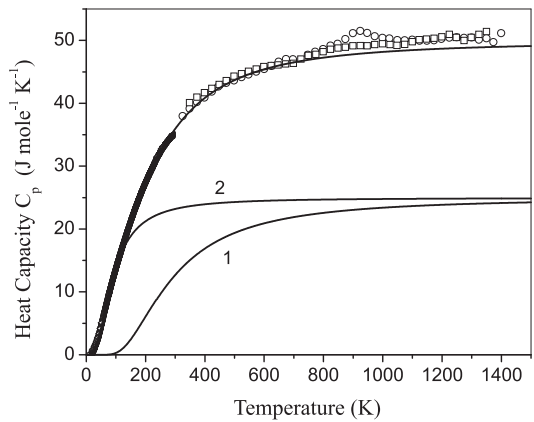}
\end{center}
\par
\centering
\caption{Dependence of heat capacity {\protect $C^{(\rm exp)}_{\rm p}$} on temperature measured by two methods: adiabatic calorimetry in the temperature range 5-70 {\rm K} and the heat-flow method in the range 20-300 {\rm K}. Thin line represents results for the dual-temperature model. Open circles and squares represent results of measurements by Kremer et al \protect \cite{4}. Thick line -- results of our measurements. The contributions of Einstein's (curve no 1) and Debye's models (curve 2) are also shown.}
\label{fig:1}
\end{figure}

In Fig. \ref{fig:1} full circles represents the total specific heat of GaN single crystal measured by heat flow method upon heating and cooling in the temperature range 20-300 {\rm K} and by adiabatic method in the range 5-70 {\rm K}(thick solid line). In the common temperature interval 20-70 K, the results obtained by both methods coincide.  

\begin{figure}[tbp]
\begin{center}
\includegraphics[width=7cm]{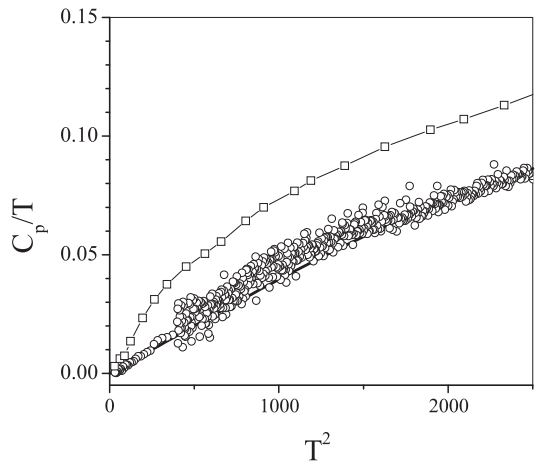}
\end{center}
\caption{{\protect $C^{(\rm exp)}_{\rm p}/T$} versus {\protect $T^2$} for hexagonal GaN in the low temperature region. The solid line presents Debye's heat capacity $C_{\rm D}$ with  $\theta_{\rm D}=365 \rm K$. The experimental points are shown as open circles. The squares represent experimental results obtained by Koshchenko et al \protect \cite{3} for powder GaN samples.}
\label{fig:3}
\end{figure}

We consider the contributions of the Debye heat capacity $C_{\rm D}$ and the Einstein heat capacity $C_{\rm E}$ to the total heat capacity $C_{\rm p}\approx C_{\rm D}+C_{\rm E}$. For temperature independent Debye's and Einstein's temperatures  $\theta_{\rm D} = 365$ ${\textrm K}$,  $\theta_{\rm E} = 880$ ${\rm K}$, the plot of the function (\ref{eq:total-c_p}) approximates the heat capacity obtained in our experiments with 3$\%$ accuracy (Fig. \ref{fig:1}). The $C_{\rm p}$ values measured below T = 7 {\rm K} are less accurate. If $C_{\rm p}$ exhibits the $T^3$ behavior, and if the free carriers contribution is negligible, then a $C_{\rm p}/T$ versus $T^2$ plot results in a straight line through the origin. This means that for the specimens studied, the electron contribution to $C_{\rm p}$ is negligible. Fig. \ref{fig:3} shows $C_{\rm p}/T$ versus $T^2$ plot in the range of 0-45 {${\rm K}$}. From this figure it can be concluded that $C_{\rm p}/T$ is proportional to $T^2$.
 
We also calculated the Debye temperature using familiar formulas \cite{13} and the room temperature elastic constants for {\rm GaN}. The results are summarized in Table \ref{table:1}. The value of $\theta_{\rm D}$ providing the best fit of $C_{\rm p}(T)$ is smaller than values collected in Table \ref{table:1}. It is interesting to note that in the case of measurements for powder samples \cite{3} $\theta_{\rm D}$ depends on temperature. 

\begin{table}
\begin{tabular}
[c]{|c|cccccccc|c|}\hline
ref. no  & 1 & 2 & 3 & 4 & 5 & 6 & 7 & 8 & {\protect \small our res.}\\
\hline
$\theta_{\rm D}$ (K) & 618 & 404 & 651 & 657 & 641 & 563 & 632 & 632 & 365\\
\hline
%\ref{table:1}
\end{tabular}
\caption{Debye's temperature for various sets of elastic constant and our result\\
{\scriptsize
$^1)$ C. Deger, E. Born, H. Angerer, O. Ambacher, M. Stutzmann, J. Hornsteiner, E. Riha, G. Fischerauer, 
Appl. Phys. Lett.72, 2400 (1998)\\
$^2)$ V. A. Savastenko, A. U. Sheleg, Phys. Stat. Sol. A 48, K135 (1978)\\
$^3)$ A. Polian , M. Grimsditch , I. Grzegory, J. Appl. Phys. 79, 3343 (1996)\\
$^4)$ Y. Takagi, M. Ahart, T. Azuhata, T. Sota, K. Suzuki, S. Nakamura, Physica B 219/220, 547 (1996)\\
$^5)$ M. Yamaguchi, T. Yagi, T. Azuhata, T. Sota, K. Suzuki, S. Chichibu, S. Nakamura, J. Phys. C 9, 241 (1997)\\
$^6)$ R. B. Schwarz, K. Khachaturyan, E. R. Weber, Appl. Phys. Lett.70, 1122 (1997)\\
$^7)$ T. Deguchi, D. Ichiryu, K. Toshikawa, K. Sekiguchi, T. Sota, R. Matsuo, T. Azuhata, T. Yagi, S. Chichibu, S. Nakamura,
J. Appl. Phys. 86, 1860 (1999)\\
$^8)$ M. Yamaguchi, T. Yagi, T. Sota, T. Deguchi, K. Shimada, S. Nakamura, J. Appl. Phys. 85 , 8502 (1999)}}
\label{table:1}
\end{table}

\begin{figure}[htpb]
\begin{center}
\includegraphics[width=7cm]{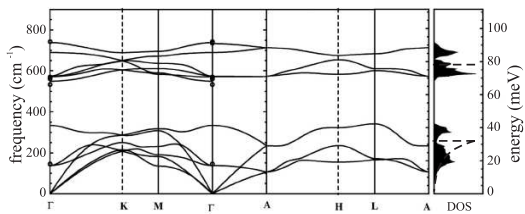}
\end{center}
\par
\centering
\caption{{\it Ab initio} phonon dispersion curves and density of states DOS for GaN in the wurtzite structure  \cite{9}. The dashed line is the Debye DOS Eq. \protect (\ref{Debye-DOS}) for $\hbar \omega_{\rm D}=32$ ${\rm meV}$. The Einstein energy \protect $\hbar \omega_{\rm E}=77$ ${\rm meV}$ is also indicated.}
\label{fig:5}
\end{figure}
The main feature of the phonon DOS of hexagonal GaN \cite{6}-\cite{9} is the presence of two well separated bands of low energy phonons (acoustic and optic) in the range 0-42 {\rm meV} and optic phonons in the range 70-90 {\rm meV} (cf. Fig. \ref{fig:3}). Therefore, analysis of our $C_{\rm p}(T)$ data  with the help of two temperatures $\theta_{\rm D}$ and $\theta_{\rm E}$ is sound. The cut-off energy $k_{\rm B}\theta_{\rm D}=32$ ${\rm meV}$ is located near the upper limit of the lower energy band. The Einstein energy $k_{\rm B}\theta_{\rm E}=77$ ${\rm meV}$ is located near the center of the energy band for optic phonons. 

Consider the kinetic expression for the thermal conductivity coefficient $\kappa$ 
\begin{equation}
	\kappa=C_{\rm p}l_{\rm ph}\bar{v}/3,
\end{equation}
where $l_{\rm ph}$ is the phonon mean free path and $\bar{v}$ is the mean velocity of acoustic phonons. Using it we estimate the experimental phonon mean free path $l_{\rm ph}$. Obviously $\bar{v}$ and $l_{\rm ph}$ are characteristics of the ``propagative" phonons, i.e., those which really contribute to the transport phenomena. Consequently the heat capacity $C_{\rm p}$ to be considered should be only due to the contribution of these phonons, i.e., $C_{\rm p}\approx C_{\rm D}$. For GaN $\bar{v}\approx5\times 10^{5} {\rm cm/s}$. Since we have established the temperature dependence of $\kappa$ \cite{2} of the specimen which was also used in our heat capacity measurements, we can find the dependence of $l_{\rm ph}$ on temperature. 
The results of such analysis are presented in Fig. \ref{fig:5}. Below 100 {\rm K} the mean free path is proportional to $T^{-2}$. Above 100 {\rm K} the Umklapp-processes gain importance \cite{14}, making mean free path shorter. For 300 {\rm K} the phonon mean free path exceeds 100 {\rm nm}, which is 10 times greater than characteristic length of nanodevices based on low dimensional gases of carriers \cite{15},\cite{16}. This means that in such devices the energy removal by acoustic phonons with energies lower than $k_{\rm B}\times 300$ ${\rm K}$ is ballistic. This observation is in agreement with our previous results for heat pulses propagating in bulk GaN \cite{17}. 
\begin{figure}[htbp]
\begin{center}
\includegraphics[width=7cm]{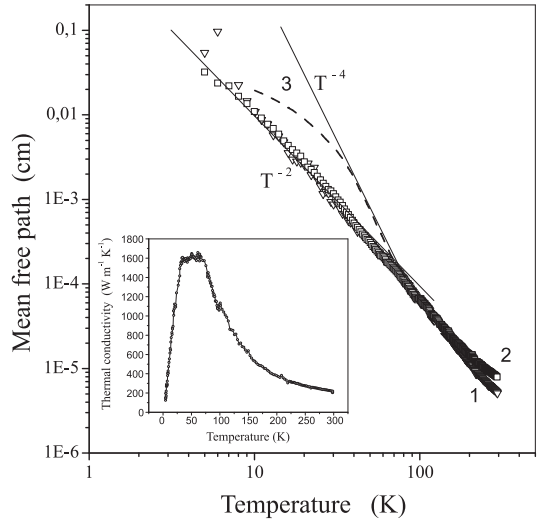}   
\end{center}
\par
\centering
\caption{The temperature dependence of the phonon mean free path for the wurzite \protect {\rm GaN} monocrystalline specimen. Triangles (line no. 1) indicate \protect $l_{\rm ph}^{(\rm exp)}=3\kappa_{\rm exp}(T)/\bar{v}C_{\rm p}^{(\rm exp)}(T)$ and squares (line no. 2)  stand for \protect $l_{\rm ph}=3\kappa_{\rm exp}(T)/\bar{v}C_{\rm D}(T)$. The broken line represents results obtained by Liu and Balandin \protect \cite{14}. Insert --  heat conductivity coefficient \protect \cite{2}.}
\label{fig:6}
\end{figure}

\section*{Acknowledgements}
%Acknowledgements
B.A. Danilchenko would like to express his gratitude for grant STCU-Project 3922.

\end{document}